# BANDWIDTH AND DELAY ISSUES ON THE NETWORK ROUTING

**Ahmad Mateen[1], Muzaffar Hussain[2] and M. Yahya Saeed[3]**
Department of Computer Sciences, University of Agriculture Faisalabad.
Corresponding Authour: ahmedmatin@hotmail.com

***ABSTRACT***— *N/W has always been intended to be incredibly strong. H/W and software changes affect basic structure on the N/W capacity. N/W-Routing-protocols have been used to determine the shortest path as the goal. N/W information is deployed using dominant interior N/W-Routing-protocols. Although in N/W little instability is associated with bandwidth and latency this requires analysis. Statistical analysis of the samples have been taken for identifying the relationship between dependent and independent variables. Study has been made to see the relations with the delay, bandwidth, and N/W-Routing stability to draw to the conclusions for a positive relationship. Thus, for getting the better performance and results, bandwidth and delay should be reduced, as suggested in this paper.*

## 1. INTRODUCTION

In past, NL has been intended for N/W-Routing traffic between N/Ws. NL is made because of many forms of communication, including telephone N/W (circuit switched), electronic data N/Ws (like Internet) and transportation N/Ws. This has been first done with electronic data communication path, using Data-Packets switching technology. In Data-Packets-switched N/Ws, N/W-Routing directs Data-Packets transfer (N/W that has been logically addressed source when they went to their ultimate) through intermediate nodes. Generally intermediate N/W machines like N/W routers, bridges, gateways, firewalls, or replaced. General purpose computers may also be Data-Packetss and NL, and they have been not specialized H/W and may also suffer from limited performance. Process route usually directs forwarding on basis of N/W-Routing tables this maintain a record of lines of different areas of N/W. So we build N/W-Routing tables, this take place in router's memory, it has been most effective way. Most N/W-Routing algorithms using only one side of NL N/W at a time. Multipath N/W-Routing strategies enable use of several different ways[1]. NL, in a sense, a little more of term, has been often compared to on assumption that N/W addresses have been structured and same implies close within an N/W. Address structured to enable an N/W-Routing table entry to represent route group N/W machines. In large N/Ws, structured addressing (N/W-Routing, narrow sense) outperforms unstructured addressing (bridging). NL is in excellent form on Internet. [2].

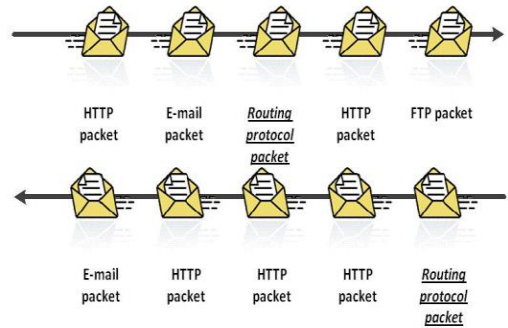

**Fig 1: Packet Flow Over a NL**

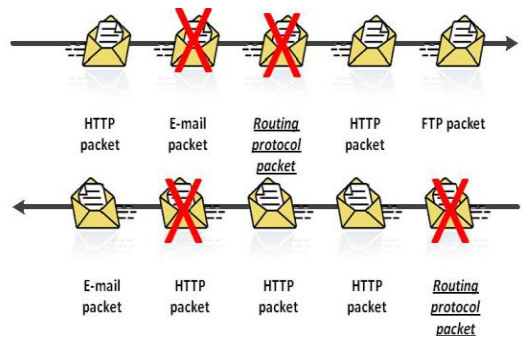

**Fig 2: Packet Discarding**

### 1.1. BANDWIDTH LIMITATION AND N/W-ROUTING-PROTOCOLS

Today, the stability of the commercial N/W is very important. Many companies rely on the cooperation between different departments and fast and reliable N/W is a must to meet strict deadlines. Any large N/W using an N/W-Routing protocol, N/W-Routing data Data-Packetss between users. N/W-Routing-protocols may induce unstable N/Ws, if the NL bandwidth between them under a restricted or poor NL quality, Data-Packets loss transit. In bandwidth constrained NLs, N/W-Routing protocol may be problems. If the N/W-Routing protocol Data-Packets exceeds the limit, they have been dropped nearby may be lost cause withdrawal N/W-N/W-routes. [6].

### 1.2. DELAY IN N/W-ROUTING

In the delay fault-tolerant N/W, attention and the ability to move your destination, or route data source, communications should be set up to the basic route. Delay and disruption tolerant N/Ws (DTNs) lack connectivity features and the approach leads to a lack end-to-end. In these challenging areas, popular ad hoc N/W-Routing-protocols like AODV and DSR may not establish N/W-N/W-routes. This is because these protocols first establish a complete route, and then the route has been selected after the actual data transfer. Somehow, as the end to end in the way, it difficult or impossible to establish the N/W-Routing protocol and it must use a "store and forward" method, where data movement phases. It would come to its place to keep the whole N/W working. The normal way to increase the probability that the message is successfully transferred many copies of the message will be reaching the destination [7].

## 1.3. INTERNET N/W-ROUTING INSTABILITY
Preparing to NL redundant Internet Company give a complete service; reduce power outages and related costs. This strategy also provides peace of mind as a bonus for the administrator. Here is how to use the Border Gateway Protocol (BGP) have the same results for your company. With VPN, e-commerce, and other important applications, the Internet has become needed for many companies mission criticaly to ensure the availability of these applications. In the grounds it helps Internet connection to corporate performance Internet access costs. BGP is one of the key tools to achieve good N/W-Routing Internet connection. If location is connected to two different versions of the Internet, it is called multi-homed. If one has multi-homed N/W for two different ISP's, BGP runs Internet router (S) on, and choosing the ISP to provide the resources to provide the best performance and N/W optimization.

## 2. PREVIOUS STUDY
Dlay and disruption tolerant N/Wing (DNT) is a very exciting research for more and more researchers. It focuses on the design, construction, evaluation and application architectures, services and protocols to achieve heterogeneous N/Ws, applications, during the connection tolerance beyond the traditional IP N/W, even some of the delay between the losses of data interoperability. DTNs may include the deployment, but not limited to, N/W Star, the military and tactical programs, N/W disaster, wildlife tracking / monitoring sensor N/Ws, Telematics and Communications in rural and remote developing countries. Chapter begins with a review of DNT architecture, its key concepts. Then, DNT situations described in the last chapter and proposed classification based N/Ws represent the DNT and N/W-Routing-protocols [8]. In the study of DTNs, while giving attention on the application of different issues to go in the context of addressing and N/W-Routing data. This study describes these existing analysis; they have the ability to make a reliable stream of data transmission scheme regardless of route N/W delay tolerant. This study sees the development of any of DNT to achieve the basic algorithms and mechanisms needed to pay close attention to discuss the delay tolerant shrubs N/W (DNT) by taking the performance-related. First, the indicators, the system of DNT performance-related may not be identified. It affects the quality of the data and signal transmission IO and performance that may be achieved have been analyzed and energy efficiency, and had discussed the tradeoff between delay. Real-world system data for the implementation of the message based on the discussion of issues and experience with the number of DNT, and recommendations will be given to the designer of the system.

Opportunistic N/W-Routing protocol recently appeared fine delay tolerant N/W N/W-Routing-protocols. In this research, it has been discussed some inspiration opportunistic N/W-Routing, and to identify a number of key issues in the design of opportunistic successfully N/W-Routing protocol. Paper reviews a number of representative opportunistic N/W-Routing-protocols, including extremely opportunistic N/W-Routing protocol, adaptive N/W-Routing protocol simple opportunity, the opportunity elastic mesh N/W-Routing, multi-code conversion, overlay performance-based, opportunistic N/W-Routing, understanding social opportunistic N/W-Routing protocol discussed in protocol , the central idea, key assumptions, progress and shortcomings. We identify opportunities to open lines for research question.

## 3. METHODOLOGY
### 3.1. THEORETICAL FRAMEWORK
The N/W-Routing is dependent variables while the bandwidth and delay have been independent variable the framework have been given below.

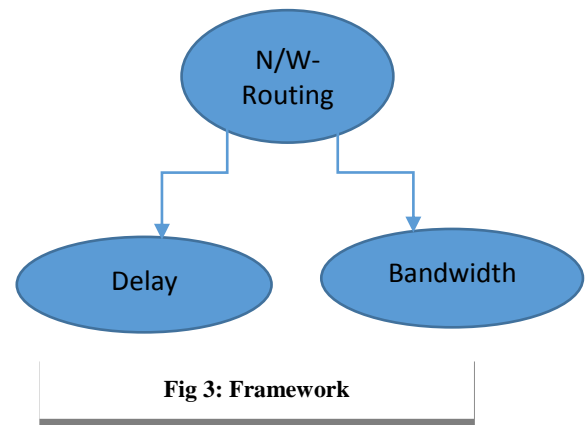

**Fig 3: Framework**

Delay and disruption tolerant N/Wing (DNT) is a very exciting research for more and more researchers. It focuses on the design, construction, evaluation and application architectures, services and protocols to achieve heterogeneous N/Ws, applications, during the connection tolerance beyond the traditional IP N/W, even some of the delay between the losses of data interoperability. DTNs includes the deployment, but not limited to, N/W Star, the military and tactical programs, N/W disaster, wildlife tracking / monitoring sensor N/Ws, Telematics and Communications in rural and remote developing countries. It reviews the DNT architecture, its key concepts as well. Then, DNT situations described in the last chapter and proposed classification based N/Ws represent the DNT and N/W-Routing-protocols.

### 3.2. ANALYSIS TOOLS (SPSS)
Although the findings provide data about the control over the organization spreadsheet program, and may be seen as some form. In contrast, you may not SPSS, because the best way to organize the data and the movement of block data. Line represents a case, while a representative of the number. If you have been using a spreadsheet, the user must manually define the analysis of this relationship. SPSS is made directly to the analysis of statistical data, so that it provides a method, a large range of graphs and charts. The overall process may offer other services, like: invoices and financial statements, but specialy this work better. In addition, the common spreadsheet may support as soon as the installation and analysis of data, need to have access to the best techniques, additional plug-ins. SPSS design must ensure that the output from the data is kept separate. In fact, it is stored in a separate file; this is different from the data of all the results. Somehow, the results of analysis programs like Excel have been placed in a worksheet, and may be accidental overwrites and other information. Although Excel data organization still provides a great way to use special software, like SPSS in-depth analysis of the data fit more [1,5].

### 3.3. SIMULATION RESULTS
N/W-Routing algorithms for Myrinet N/W simulation show the following results.
First RSUD, RRSUD and PUD follow the analysis of the impact. Later set-transit mechanism by RMIT, RRMIT and RRMIT MIN, associated to assess the benefits. Different sizes have been used, but for all the results formed with the 512-byte Data-Packets to onward.

### 3.4. UP/DOWN N/W-ROUTING WITH ALTERNATIVE N/W-ROUTES
In this section RSUD, RRSUD and PUD used to select various resolutions submitted by using the up and down direction to evaluate the behavior of the algorithm.

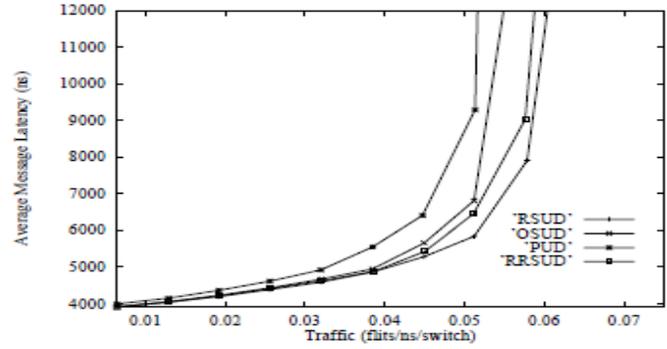

**Fig 4: Average message latency vs. traffic. Network size is 16 switches**

In addition to the low cost of TRAFFIC, different methods used to achieve the delay in the lengthier N/W-routes from algorithm and found almost the same performance.

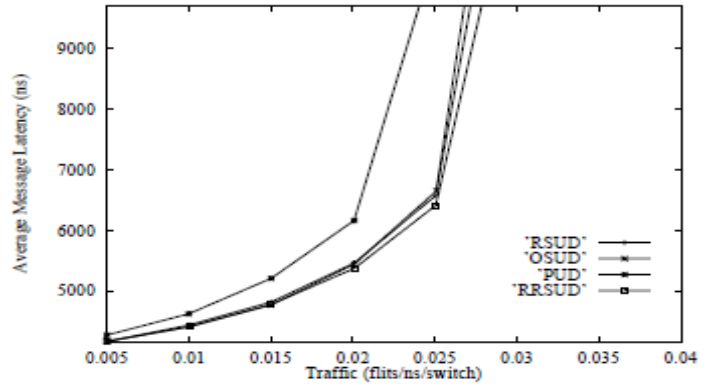

**Fig 5: Average message latency vs. traffic. Network size is 32 switches**

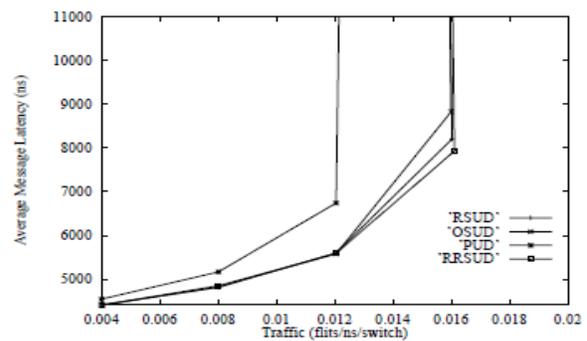

**Fig 6: Average message latency vs. traffic. Network size is 64 switches**

### 3.5. MINIMAL N/W-ROUTING WITH IN-TRANSIT
### 3.5.1. BUFFERS

Algorithm uses OUT RMIT, RRMIT and RRMIT MIN for comparison of selection methods. All the keys of sixteen and sixty-four are used to vary the size of the system. The 512-byte Data-Packets is used for the results of the 16-switch system. It shows minimal access path (based RMIT, RRMIT, and RRMIT-MIN).

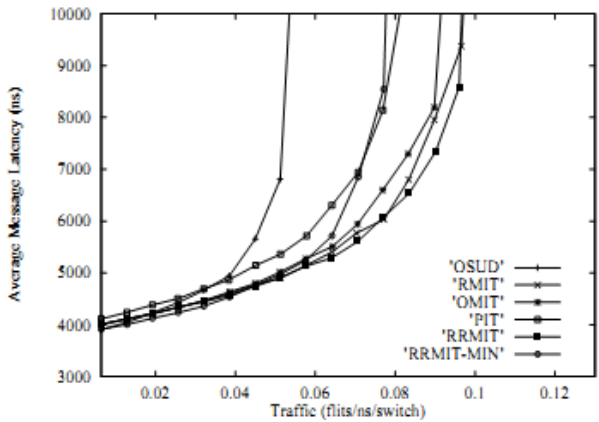

**Fig 7: Latency vs. traffic for average message and the network size is 16 switches**

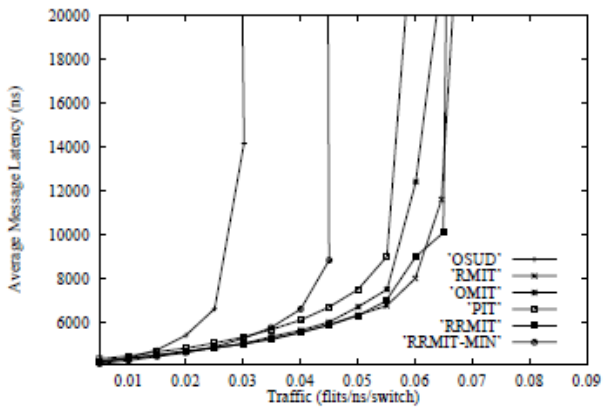

**Fig. 8: Latency vs traffic for average message and the network size is 32 switches**

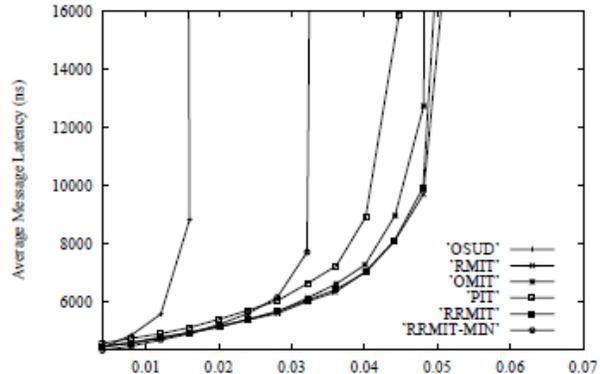

**Fig. 9: Latency vs traffic for average message and the network size is 64 switches**

This is the result of buffers associated to the RRMIT-MIN algorithm avoids this issue by minimizing using in-transit hosts.

### 3.6. CORRELATION ANALYSIS

Data processed from SPSS. The correlation analysis of values is given as under.

**Table 1: Correlations**

Correlations

| | | Routing Values | Bandwidth Values | Delay Values |
|---|---|---|---|---|
| Routing Values | Pearson Correlation | 1 | .163 | .187 |
| | Sig. (2-tailed) | | .505 | .562 |
| | N | 15 | 15 | 15 |
| Bandwidth Values | Pearson Correlation | .163 | 1 | .183 |
| | Sig. (2-tailed) | .505 | | .714 |
| | N | 15 | 15 | 15 |
| Delay Values | Pearson Correlation | .187 | .183 | 1 |
| | Sig. (2-tailed) | .562 | .714 | |
| | N | 15 | 15 | 15 |

This correlation table shows that the relationship between the bandwidth and N/W-Routing is very strong if the bandwidth increases results in that the N/W-Routing is stable. Somehow if the bandwidth is decreased N/W-Routing affects the internet speed and browsing.

### 3.7. REGRESSION ANALYSIS
The regression analysis is below:

**Table 2: Regression Summary Model**

Variables Entered/Removed[a]

| Model | Variables Entered | Variables Removed | Method |
|---|---|---|---|
| 1 | Delay Values, Bandwidth Values[b] | . | Enter |

a. Dependent Variable: Routing Values
b. All requested variables entered.

**Table 3: Regression**

Model Summary

| Model | R | R Square | Adjusted R Square | Std. Error of the Estimate |
|---|---|---|---|---|
| 1 | .236[a] | .056 | .102 | 11.18410 |

a. Predictors: (Constant), Delay Values, Bandwidth Values

In model summary the value of R is 0.236 and it's very strong. It's show the exact relation between the variables.

### 3.8. ANOVA

**Table 4: ANOVA**

| Model | | Sum of Squares | df | Mean Square | F | Sig. |
|---|---|---|---|---|---|---|
| 1 | Regression | 88.723 | 2 | 44.362 | .355 | .709[b] |
| | Residual | 1501.010 | 12 | 125.084 | | |
| | Total | 1589.733 | 14 | | | |

a. Dependent Variable: Routing Values
b. Predictors: (Constant), Delay Values, Bandwidth Values

In above table shows that it is very substantial variability in the dependent variable from variability in the independent variables and the residual (sum of squares) is 1501.010. The total (sum of squares) is 1589.733. The F-statistic is .355, it show the consequences of the variables.

### 3.9. COEFFICIENTS

**Table 5: Coefficients**

Coefficients[a]

| Model | | Unstandardized Coefficients | | Standardized Coefficients | t | Sig. |
|---|---|---|---|---|---|---|
| | | B | Std. Error | Beta | | |
| 1 | (Constant) | 10.819 | 4.873 | | 2.220 | .046 |
| | Bandwidth Values | .080 | .131 | .172 | .610 | .553 |
| | Delay Values | .123 | .239 | .145 | .515 | .616 |

a. Dependent Variable: Routing Values

In Coefficient table, we see the very less difference between the variables. The slope of bandwidth is -.080 and the value of delay is .123. Its show the reverse relation between the N/W-Routing stability and delay. If we have increased the N/W-Routing then the delay will be decreased

## 4. CONCLUSION
It is concluded that the relationship of dependent variable and independent variable are not the same. The relationship between the N/W-Routing stability and delay is reversed i.e increase in the delay means decrease in routers stability and its definatily affect your browsing and working. So, increase your bandwith with our proposed path selection algorithm in order to improve the efficiency and better stability of router's, which improves the over all N/W performance.